\PassOptionsToPackage{pagebackref,breaklinks,colorlinks,allcolors=cvprblue}{hyperref}

\documentclass[10pt,twocolumn,letterpaper]{article}

\usepackage{cvpr}              %
\usepackage{soul}     %
\usepackage{url}
\usepackage{tabularx}

\usepackage{graphicx}
\usepackage{booktabs}
\usepackage{blindtext}%
\usepackage{graphicx}
\usepackage{bm}
\usepackage{multirow}
\usepackage{pifont}%
\usepackage{array}
\usepackage{colortbl}
\usepackage[flushleft]{threeparttable}
\usepackage{comment}
\usepackage{float}
\usepackage{balance}
\usepackage[normalem]{ulem}
\usepackage{array}
\usepackage{wrapfig}
\usepackage{capt-of} 
\usepackage{xcolor}   %

\usepackage{natbib}
\usepackage{xurl} %
\usepackage{orcidlink}

\definecolor{SoftBlue}{RGB}{224,240,255}
\definecolor{SoftPurple}{RGB}{242,232,246}
\definecolor{SoftGold}{RGB}{255,244,214}
\definecolor{rebutviolet}{RGB}{128, 0, 200}       %

\definecolor{noteteal}{RGB}{0,128,128}

\definecolor{rebutorange}{RGB}{230, 120, 20} %

\newcommand{\mesh}{M}
\newcommand{\shape}{\bm{\beta}}
\newcommand{\albedo}{\bm{\alpha}}
\newcommand{\pose}{\bm{\theta}}
\newcommand{\expression}{\bm{\psi}}

\newcommand{\smplx}{\mbox{SMPL-X}\xspace}

\renewcommand{\paragraph}[1]{\noindent\textbf{#1}}

\definecolor{cvprblue}{rgb}{0.21,0.49,0.74}

\def\paperID{*****} %
\def\confName{CVPR}
\def\confYear{2026}

\title{Evaluation of Generative Models for Emotional 3D Animation Generation in VR}

\author{Kiran Chhatre\qquad
Renan Guarese\qquad
Andrii Matviienko\qquad
Christopher Peters\\
KTH Royal Institute of Technology\\
\href{https://emotional3dhumans.github.io/}{emotional3dhumans.github.io}
}

\begin{document}

\twocolumn[{%
\maketitle
\vspace{-2.5em}
\begin{center}
\includegraphics[width=\textwidth]{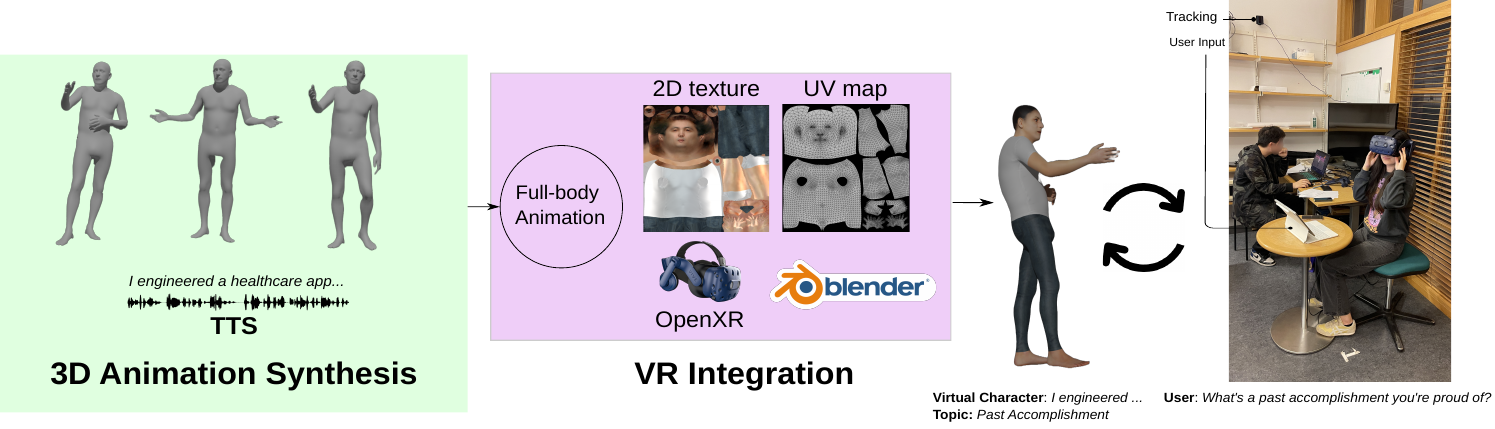}\par
\vspace{-0.4em}
{\captionsetup{hypcap=false}
\captionof{figure}{\textbf{Evaluation of Generative Models for Emotional 3D Animation in VR.}
In this evaluation, participants interact with a virtual character using a VR headset. The setup is modular and supports integration of various text-to-speech (TTS) models and speech-driven 3D animation generation methods. On the right, the figure illustrates an interaction between the participant and the virtual character. Participants' positions are tracked by two base stations installed in the study room, and they use a tablet to record input during the session. The animation generation method utilizes speech segments generated by a TTS system to produce corresponding 3D facial expressions and body animations. These predicted animation data are mapped onto a 3D character, textures are applied via UV mapping, and the final content is rendered and streamed in real-time for VR interaction using Blender (OpenXR).}
\label{fig:system}}
\end{center}

}]

\begin{abstract}

Social interactions incorporate various nonverbal signals to convey emotions alongside speech, including facial expressions and body gestures. Generative models have demonstrated promising results in creating full-body nonverbal animations synchronized with speech; however, evaluations using statistical metrics in 2D settings fail to fully capture user-perceived emotions, limiting our understanding of the effectiveness of these models. To address this, we evaluate emotional 3D animation generative models within an immersive Virtual Reality (VR) environment, emphasizing user-centric metrics—emotional arousal realism, naturalness, enjoyment, diversity, and interaction quality—in a real-time human–agent interaction scenario. Through a user study ($N=48$), we systematically examine perceived emotional quality for three state-of-the-art speech-driven 3D animation methods across two specific emotions: happiness (high arousal) and neutral (mid arousal). Additionally, we compare these generative models against real human expressions obtained via a reconstruction-based method to assess both their strengths and limitations and how closely they replicate real human facial and body expressions. Our results demonstrate that methods explicitly modeling emotions lead to higher recognition accuracy compared to those focusing solely on speech-driven synchrony. Users rated the realism and naturalness of happy animations significantly higher than those of neutral animations, highlighting the limitations of current generative models in handling subtle emotional states. Generative models underperformed compared to reconstruction-based methods in facial expression quality, and all methods received relatively low ratings for animation enjoyment and interaction quality, emphasizing the importance of incorporating user-centric evaluations into generative model development. Finally, participants positively recognized animation diversity across all generative models.

\end{abstract}

\section{Introduction}
Conversational interactions between users and virtual characters are crucial for immersive social experiences in VR, requiring the generation of behaviors such as speech \citep{vasiliu2025towards}, gestures \citep{ghorbani2022zeroeggs}, and facial expressions \citep{OhKruzic2020FacialEC}. However, accurately replicating verbal and non-verbal cues remains challenging. Social interactions incorporate multiple non-verbal modalities—such as gesture arousal, facial expressions, eye contact, and body posture—that are vital for conveying emotions, often taking precedence over verbal language~\citep{10.3389/fpsyg.2019.02063,communicativefacial}. Moreover, non-verbal expressions guide human behavior by providing key signals on how to respond to others~\citep{nonverbalguilt} and shape perceptions of personality traits~\citep{TRACY201525}. Yet, misconceptions persist about how these non-verbal cues function in real conversations, making it difficult to confirm whether the generated animation in virtual characters truly conveys the intended emotional behavior. These challenges highlight the need for comprehensive deep learning models that account for the interplay among multiple modalities~\citep{doi:10.1177/17456916221148142}.

In virtual environments, verbal and non-verbal expressions are essential for delivering immersive social experiences, contributing significantly to users’ social presence and emotional engagement~\citep{10.1145/3173574.3173863}. The intricate interplay of verbal and non-verbal cues complicates both the modeling and evaluation of character behavior. Early research employed hand-crafted animation and rule-based models~\citep{Cassell2001BEATTB,Poggi2005}, but these methods cannot capture the full range of possible cues, limiting the fidelity of social interactions. More recent work has leveraged motion capture to create high-fidelity behavior for teleoperated (or Wizard of Oz; WOZ) avatars~\citep{ExpressivenessFrontier, 10316418}, which excel at conveying emotional, full-body expressions by relying on human performers. Yet, this approach is costly and less scalable due to expensive motion capture technology. With the rapid development and widespread use in generating speech and motion content, generative models offer new possibilities for creating human-like social agents. Tools such as text-to-speech (TTS) systems~\citep{Kim2021ConditionalVA,DBLP:journals/corr/abs-2112-02418,vasiliu2025towards} and speech-to-animation models~\citep{yi2023generating} have opened new avenues for building the virtual characters. By leveraging these models, one can automate their creation: TTS produces natural-sounding speech for dialogue scripts, while speech-to-animation models synchronize gestures and facial expressions with spoken words, adding emotional depth to interactions~\citep{chhatre2023emotional, EMOTE}. Although recent studies demonstrate high performance in animation realism, expressiveness, and diversity in  \textit{monologue} scenarios~\citep{fan2022faceformer, chhatre2023emotional, EMOTE}, the effectiveness of these models in VR \textit{dialogue} settings—where human interactions with virtual characters come into play—remains uncertain.

Existing generative models can produce full-body animations from speech~\citep{chhatre2023emotional,danecek2023emotional,ginosar2019gestures,alexanderson2023listen,yang2023QPGesture,yang2023DiffuseStyleGesture} and provide holistic co-speech, full-body datasets~\citep{Mughal_2024_CVPR,Liu_2024_CVPR}. However, gesture generation has largely relied on \textit{objective metrics}—for instance, Frechet Gesture Distance~\citep{https://doi.org/10.48550/arxiv.2204.12318, Yoon2020Speech} (comparing latent features between generated and ground-truth motion), beat alignment~\citep{li2021aichoreographermusicconditioned, Valle_P_rez_2021} (assessing motion-speech correlation via kinematic and audio beats), semantic-relevance gesture recall~\citep{liu2022beat}, and gesture diversity~\citep{li2023audio2gesturesgeneratingdiversegestures, liu2022disco} (covering beat, deictic, iconic, and metaphoric gestures). While these metrics are useful, they often fail to capture how humans truly perceive gestures. \textit{User-centered metrics}—including perceived emotional realism, naturalness, and diversity—remain underexplored, even though they are crucial for evaluating virtual characters in a social context~\cite{10.1145/3722564.3728374}. The effectiveness of these models depends on how well users perceive expressed emotions and interactional effects during social exchanges.

Although some studies have evaluated virtual faces and gesture generation—for example, investigating the uncanny valley effect~\citep{DINATALE2023100288}, the GENEA Challenge~\citep{10.1145/3577190.3616120} on speech-driven gesture generation in monadic and dyadic contexts, research on the relationship between empathy and facial-based emotion simulation in VR~\citep{10.1145/3656650.3656691}, and AV-Flow~\citep{chatziagapi2025avflow} for dyadic speech and talking-head generation—these typically focus on either gestures or virtual faces, rather than a holistic 3D perceptual experience combining both face and body. \cite{10.1145/3722564.3728374} examine how integrating facial expressions with body gestures influences animation congruency and synchronize the generated motion with the driving speech; however, they do not address diverse or emotionally rich conversational contexts. Closer to our work,~\cite{deichler2024gesture} evaluated animations generated by such models from a third-person viewpoint in monologue or dialogue, emphasizing the impact of an immersive VR environment relative to a 2D setting. However, their study did not address real-time human–virtual character interaction or emotional conversation contexts, leaving the effects of integrating face and body unclear. Consequently, subjective qualities remain insufficiently studied for a human–virtual character in dyadic emotional interaction.

In this work, we address this gap by evaluating generative models for animation in VR, focusing on immersive human–virtual human interactions within emotionally contextual dialogues. Our user study employs two arousal conditions—Happy (high arousal) and Neutral (mid arousal)—based on the circumplex model of affect~\citep{RussellCircumplex}. Focusing on these two states allows us to examine five key perceptual factors—realism, naturalness, enjoyment, diversity, and interaction quality (see~\cref{eval})—without introducing excessive complexity into the study design. We additionally compare participant ratings with outputs from a pretrained deep-learning classifier trained on Ekman’s eight basic emotions~\citep{ekman1993}, collapsing its predictions into the same two categories (happy vs. neutral) for consistency. Limiting the scope to happy and neutral makes the experiment tractable while establishing a foundation for future studies that may explore a broader range of affective states, including complex emotions such as guilt and embarrassment, as well as negative emotions such as anger and disgust. Our specific goal is to assess the perceptual impact of these two emotional conditions and, based on the findings, iteratively improve the study to better capture the perceived effects of more diverse affective states. Building on previous work~\citep{HEESEN2024110663,liu2024emotiondetectionbodygesture,Tan2009INTEGRATINGF,CONLEY20181059,10.3389/fpsyg.2023.1199537}, we concentrate on critical perceived emotional animation attributes. We assess these qualities through a VR-based user study ($N=48$), designed to explore immersion and social presence during interaction with virtual characters, in contrast to 2D videos~\citep{10.1145/3613905.3650773}. Advances in interactive media now support qualitatively richer experiences; immersive VR, in particular, can influence users’ physiology, psychology, behaviour, and social responses~\citep{MeasuringPresence}. To investigate how high-quality, computer-generated speech-driven animations of virtual characters affect factors such as enjoyment, persuasion, and social relationship, we therefore conduct our study in a VR setting. The same methodology could later be adapted to mixed or augmented reality environments. This study highlights the importance of perceptual evaluation, as objective metrics alone cannot fully capture the validity of generated gestures. Moreover, integrating multiple generative models for real-time interaction—combining speech and 3D animation—offers a promising direction for computational interaction systems. Rather than exclusively training or refining new models, our approach emphasizes a holistic perceptual assessment of current models to guide future model development.

While numerous speech-driven face-expression and body-gesture generative models exist, we specifically chose three representative methods based on their state-of-the-art performance on objective metrics such as realism, diversity, Frechet Gesture Distance, and beat alignment, as reported by their original authors. In our implementation, we use the SMPL-X~\citep{SMPL-X:2019} parametric model for representing virtual humans in 3D. The three chosen models—EMAGE~\citep{Liu_2024_CVPR}, TalkSHOW~\citep{yi2023generating}, and AMUSE~\citep{chhatre2023emotional}—exhibited top performance, with AMUSE uniquely focusing on emotional 3D body gestures. To incorporate face animation with AMUSE, we employed FaceFormer~\citep{fan2021faceformer}, a SMPL-X–compatible, speech-driven face-expression model. Additionally, we compared these generative models to real human expressions by capturing a human performer’s 3D face and body via the PIXIE~\citep{PIXIE:2021} frame-level reconstruction method. The core of our study is a user-based evaluation that reveals the strengths and weaknesses of these generative approaches, as well as how they shape user perception in VR. By also comparing the generative outputs to reconstruction-based real expressions, we shed light on how closely current generative methods can replicate real human body and facial expressions. This evaluation informs the selection of models best suited for specific applications, depending on which attributes—such as realism, naturalness, enjoyment, diversity, or interaction quality—are most critical.  Our main contributions are as follows:

\begin{itemize}
\item To the best of our knowledge, we present the first perceptual evaluation of generative models for emotional 3D animation in real-time human–virtual character interactions within an immersive VR environment.

\item We conduct a VR-based user study ($N=48$), evaluating three representative generative methods with demonstrated capabilities in emotional animation generation.

\item We evaluate the realism, naturalness, enjoyment, diversity, and interaction quality of the generated animations and investigate their impact on user perception.
\end{itemize}

In~\cref{sec:rl}, we review related work, and in~\cref{exp}, we cover key concepts and provide an overview of our implementation details. \Cref{eval} details our user study, while \cref{res} presents the results. \Cref{dis} discusses the findings and practical implications, followed by \cref{lim}, which addresses limitations and future work. Finally,~\cref{conc} concludes the paper.

\section{Related Work}
\label{sec:rl}

\subsection{Social Interaction}

Social interaction is a complex interplay of language, gestures, and other nonverbal behaviors. The theory of embodied cognition suggests that spoken language evolved from motor actions, with empirical studies showing motor system involvement in both language and gesture production and comprehension~\citep{gentilucci2006repetitive,rizzolatti1998language}. Research indicates that gestures and spoken language function in sync during face-to-face communication, with symbolic gestures sometimes replacing verbal components~\citep{andric2013brain}. This synchronization reflects the interaction between the sensory-motor and language processing systems~\citep{bernardis2006speech,mcneill1992hand}. Nonverbal behaviors—facial expressions, gestures, posture, and gaze—are essential for conveying intentions, often enhancing or replacing verbal communication to produce a more accurate display of emotions than any single channel alone~\citep{gunter2004communicating,zhao2018transcranial}. Gestures, in particular, are tightly integrated with speech~\citep{ozyurek2014hearing,he2018spatial}. However, the intricate ways these modalities interact remain not fully understood, even in human studies, making it challenging to develop virtual agents that accurately replicate such interactions.

\subsubsection{Emotions in Social Interaction}
\label{subsubsec:emo_in_socint}

Emotion has been a central theme in social-interaction research for decades.~\cite{langgestemo} argue that the human Mirror Mechanism associates language in shared sensorimotor representations, tightly coupling gestures, speech and affect;~\cite{HUANG2023100447} show that co-regulation of such social–emotional exchanges is critical for effective collaborative learning; and~\cite{cog_tech} treat emotions as dynamic, context-dependent processes, comparing the competence of humans with that of emotionally aware artificial agents.

\begin{figure}[t]
    \centering
    \includegraphics[width=\columnwidth]{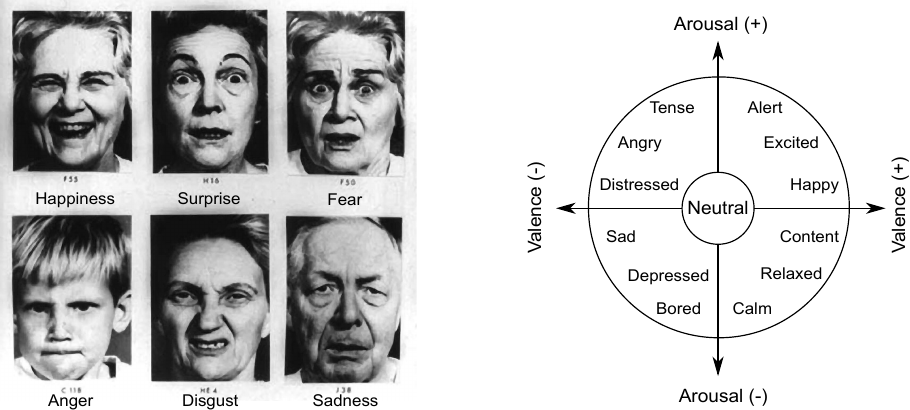}
    \caption{
      \textbf{Emotion classification.}
      Left: Ekman’s discrete-emotion theory identifies six basic categories—anger, disgust, fear, happiness, sadness, and surprise—treating each as a distinct class rather than points on a continuum~\citep{ekman1993}.
      Right: The circumplex model~\citep{RussellCircumplex} places emotions in a two-dimensional space spanned by arousal and valence; the center represents neutral arousal and neutral valence.
    }
    \label{fig:emoclass}

\vspace{-1.5em}
\end{figure}

Emotions encountered in these interactions can be cast either as discrete categories (\cref{fig:emoclass}-left) or as points in a continuous affective space (\cref{fig:emoclass}-right). Ekman’s taxonomy lists six basic classes—anger, disgust, fear, happiness, sadness and surprise—assumed to be biologically hard-wired~\citep{ekman1993}. Dimensional models assign emotions in low-dimensional manifolds:~\cite{Schlosberg1954-SCHTDO-71} organised facial expressions along pleasant–unpleasant and attention–rejection axes with activation as a third dimension, while the widely used circumplex model maps emotions onto arousal and valence axes whose origin denotes neutrality~\citep{RussellCircumplex}. Later work in the Vector model confirms the emotions are structured in terms of arousal and valence such that a positive valence represents appetitive motivation and negative valence represents defensive motivation~\citep{Bradley1992RememberingPP}. The stability of positive versus negative affect in two separate systems is analyzed into the positive activation – negative activation model~\citep {Watson1985TowardAC}. Finally,~\cite{plutchik2001nature} integrates categorical and dimensional views within a 3D framework where it arranges emotions in concentric circles, where inner circles are more basic and the outer circles are also formed by blending the inner circle emotions. Our study adopts the circumplex model, distinguishing mid- (neutral) and high-arousal (happy) conditions, and augments participant judgements with an automatic emotion recognition deep learning model trained on an extended Ekman-style eight-class taxonomy, including contempt and neutral as additional categories.

\subsection{VR-Based Interaction} %

In VR, interactions with virtual characters must be highly realistic to feel lifelike, a requirement with broad applications in entertainment and psychological research~\citep{10316418}. Approaches to creating these interactions typically fall into two categories. First, \textit{rule-based models} rely on predefined rules or human interaction knowledge~\citep{10.1007/11821830_17,Cassell2001BEATTB,Poggi2005}, often using pre-recorded animations triggered by algorithms or manual intervention~\citep{inproceedingssmartbody,10.1145/2485895.2485900,Pan2018WhyAH}. These methods are constrained by limited motion variety, leading to repetitive behaviors~\citep{10316418}. Second, \textit{teleoperation} (WOZ avatar approach) assigns human actors to drive virtual characters’ voice and body movements~\citep{ExpressivenessFrontier,Brandsttter2023DialoguesFO,10316418}. Though highly realistic, this approach depends on expensive motion capture devices and restricts the number of actors who can simultaneously participate in a single VR experience. Some studies have explored using one human to control multiple virtual characters~\citep{Osimo2015ConversationsBS,Yin2022TheOS, Brandsttter2023DialoguesFO, Yin2024WithOW}, but this reduces the variety of generated behaviors~\citep{Yin2022TheOS}, limiting scalability for group interactions in VR.

Recently, industry applications have emerged for generating narrated avatar videos~\citep{hedra,synthesia,mesh,soulmachines}, 3D interactive non-player characters~\citep{inworld,convai,ace}, and user-interactable virtual characters~\citep{replika}. However, many of these platforms lack flexibility and seamless integration with tools like Blender or Unity~\citep{blender,ue,unity}, hindering direct comparison with rule-based or teleoperated methods.

Despite these challenges, numerous studies have examined conversational virtual characters in VR~\citep{Smith2018CommunicationBI,9756815,Herrera2018EffectOB}, focusing on aspects like rendering realism~\citep{Kokkinara2015AnimationRA, Zibrek2018TheEO, Patotskaya2023AvoidingVH}, animation realism~\citep{Guadagno2007VirtualHA, doesntmatter}, facial expressions and eye gaze~\citep{Roth2018BeyondRA, Roth2018EffectsOH}, body gestures~\citep{10.1007/978-3-030-85613-7_11}, subtle social cues~\citep{MediaEq}, and emotion disclosure~\citep{PsychologicalBarreda,10.1145/1240624.1240764}. Yet most rely on rule-based or teleoperated animations, limiting both variety and quality of generated behaviors.

\subsection{Generative Models for Virtual Character Interaction}
\label{RQs}

Generative probabilistic models are widely used to produce speech and human motion. Recent advances in conditional constraints enable virtual social interactions with specific styles or emotions, offering low-cost, automated generation and diverse behaviors due to their probabilistic nature~\citep{11015786}.

Recent methods employ deep neural networks to create motion animations, emphasizing convincing non-verbal behaviors. They generate 3D talking heads from speech~\citep{pham2017end, pham2017speech, karras2017audio, taylor2017generalized, zhou2018visemenet, cudeiro2019capture, richard2021meshtalk, fan2022faceformer, xing2023codetalker} and synthesize 3D body gestures~\citep{ginosar2019gestures, qi2023diverse, Yoon2020Speech, Habibie2022AMM, yang2023QPGesture}. Some jointly produce body and facial animations via \smplx~\citep{SMPL-X:2019, yi2023generating}, enabling more expressive behaviors. While speech-driven animation control remains underexplored, recent studies introduce motion style control~\citep{yin2023emog,alexanderson2023listen} and include style and emotion constraints~\citep{fan2022faceformer,chhatre2023emotional}. 

\begin{table*}[t]
\centering
\small
\setlength{\tabcolsep}{6pt}
\renewcommand{\arraystretch}{1.15}
\caption{\textbf{Comparison of methods for 3D animation generation.}
We compare three generative models for gesture animation in VR. Methods differ in (i) generative architecture for facial expressions and body gestures, and (ii) audio processing; one also uses transcripts as extra input. All methods output body pose ($\pose$) and facial expression ($\expression$) parameters.}
\label{tab:comparison_methods}

\begin{tabularx}{\textwidth}{@{}l X X c@{}}
\toprule
\textbf{Method} & \textbf{Generative model type} & \textbf{Audio model type} & \textbf{Text} \\
\midrule
EMAGE & VQ-VAE & Content-, rhythm-aware TCN & \checkmark \\
TalkSHOW & Transformer (face), VQ-VAE (body) & Wav2Vec 2.0 & \texttimes \\
AMUSE + FaceFormer & Transformer (face), diffusion (body) & Content-, emotion-, style-aware ViT & \texttimes \\
\bottomrule
\end{tabularx}

\end{table*}

For speech generation, text-to-speech (TTS) systems allow emotional variation in tone, pitch, and rhythm~\citep{Kim2021ConditionalVA,DBLP:journals/corr/abs-2112-02418}, thereby enhancing user engagement in virtual interactions. Although individual models for speech and animation show promise, they are often developed and evaluated in isolation. In contrast, our approach integrates TTS and generative animation into a unified VR system, enabling a more comprehensive evaluation. We specifically examine how effectively they convey user perception of 3D full-body emotional responses and how these factors impact interaction quality in immersive environments.

\section{Implementation Details}
\label{exp}

\subsection{Preliminaries: Geometry, Appearance, and Rendering}\label{subsec:prelim}

We adopt the \smplx model~\citep{SMPL-X:2019} to represent 3D body geometry, defined by \(\mesh(\shape, \pose, \expression)\). This model generates a mesh \(\mesh\) from the identity shape \(\shape \in \mathbb{R}^{300}\), pose \(\pose \in \mathbb{R}^{J \times 3}\), and facial expression \(\expression \in \mathbb{R}^{100}\), where \(J\) represents the number of body joints. For its appearance, we use \smplx UV coordinates, and the shaded textures are obtained by sampling albedo \(\bm{\alpha}\), surface normals, and lighting. The Embodied Conversational Agent~\citep{10.1145/332051.332075} SMPL-X meshes—referred to as the ``agent'' hereafter—are animated in Blender using outputs from the generative models summarized in~\cref{tab:comparison_methods} and detailed in~\cref{subsec:genmodels}.

\subsection{Generative Models}\label{subsec:genmodels}

As shown in~\cref{fig:system}, we fully synthesize a virtual character’s motion and speech. We select three state-of-the-art models based on their performance in generating synthetic animations driven solely by audio input. These audio-based models generate 3D motion from speech and transcripts, and each has demonstrated strong speech-driven animation capabilities. In our pipeline, the driving speech—or video for the reconstruction baseline—is first fed to the selected model to predict full-body animation parameters. The resulting motion is then retargeted to a textured SMPL-X agent and placed in an outdoor Blender scene with appropriate lighting and camera placement. Finally, the animated scene is streamed to participants in real-time conversation through an HTC Vive Pro 2 headset. We conduct quantitative evaluations comparing all models. Each method is applied to predefined scenarios with unique topics; transcripts and speech are generated via TTS, which then drive the 3D motion. The system is modular, allowing any component to be replaced as needed.

We utilize three state-of-the-art audio-driven generation models compatible with the \smplx mesh: EMAGE~\citep{Liu_2024_CVPR}, TalkSHOW~\citep{yi2023generating}, and a combination of AMUSE~\citep{chhatre2023emotional} (for body) and FaceFormer~\citep{fan2022faceformer} (for face). In~\cref{tab:comparison_methods}, we summarize the specifics of each model. All models take raw audio as input and produce 3D animations. EMAGE and TalkSHOW output both $\expression$ and $\pose$ parameters, whereas AMUSE outputs $\pose$ parameters and FaceFormer outputs $\expression$ parameters; both parameter sets are integrated at the frame level after inference. %
Specifically, FaceFormer outputs meshes with the FLAME topology \citep{FLAME:SiggraphAsia2017}. 
We convert these meshes into FLAME expression parameters by fitting the registered 3D mesh to the FLAME model using the FLAME fitting framework~\citep{tf_flame_2022} and the Broyden-Fletcher-Goldfarb-Shanno optimizer. 
Once we obtain the $\expression$ parameters, we combine them with the $\pose$ parameters—aligning jaw rotations framewise—to create a single motion file. 
Throughout this process, the identity parameters ($\shape$) from the original AMUSE output are preserved.%
Next, EMAGE accepts text transcripts as an additional input. All geometric parameters are passed to the SMPL-X Blender add-on, which imports the meshes into the Blender scene. Each imported SMPL-X mesh includes a shape-specific rig and blend shapes for shape, expression, and pose parameters. We use consistent sampled $\shape$ parameters and an $\albedo$ texture across all models. All evaluated models—EMAGE~\citep{Liu_2024_CVPR}, TalkSHOW~\citep{yi2023generating}, AMUSE~\citep{chhatre2023emotional}, and FaceFormer~\citep{fan2021faceformer}—were made publicly available by their respective authors. An introduction to each method is provided in the Supplementary Material (Sec. 2).

\begin{figure*}[t]
    \centering
    \includegraphics[width=\textwidth]{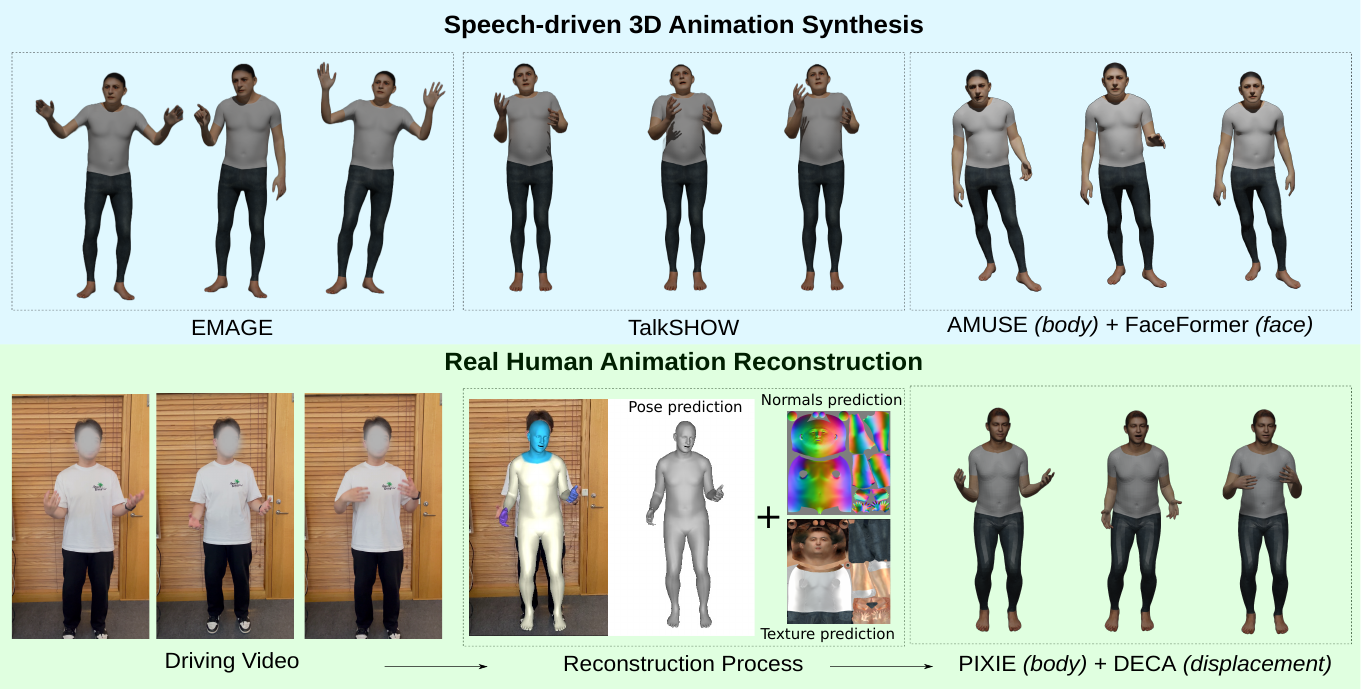}
    \caption{
      \textbf{Qualitative evaluation.}
      Top: Specific frames from the generated animation sequences using EMAGE~\citep{Liu_2024_CVPR}, TalkSHOW~\citep{yi2023generating}, and a combination of AMUSE (body animation)~\citep{chhatre2023emotional} and FaceFormer (facial expressions)~\citep{fan2021faceformer}.
      Bottom: The workflow for generating reconstruction-based animations from real human facial expressions and body gestures using driving video input, which serves as our baseline. The reconstruction method PIXIE~\citep{PIXIE:2021} + DECA~\citep{DECA:Siggraph2021} predicts pose parameters, normal maps, and textures, which are combined and rendered. Specific frames from the resulting video-based reconstruction animations are shown in the bottom right.
    }
    \label{fig:objanim}
    
\vspace{-1.0em}
\end{figure*}

The models process audio features differently. TalkSHOW uses a pre-trained Wav2Vec~\citep{NEURIPS2020_92d1e1eb} model to extract speech features, while EMAGE and AMUSE employ specialized models for this purpose. EMAGE uses a content- and rhythm-aware Temporal Convolutional Network (TCN)~\citep{Lea_2017_CVPR} that distinguishes gestures related to semantic content versus rhythm for each frame. FaceFormer also uses Wav2Vec to extract speech features, whereas AMUSE uses a Vision Transformer (ViT)-based model~\citep{dosovitskiy2020image}. The AMUSE model additionally disentangles content-, \mbox{emotion-,} and style-aware features from the driving speech, explicitly modeling the impact of emotions on generated gestures. The backbone architectures used for gesture and expression generation vary among the models. EMAGE utilizes multiple Vector Quantized Variational AutoEncoders (VQ-VAE)~\citep{van2017neural} to generate both facial and body animations. TalkSHOW employs a VQ-VAE for body animation, while a standard encoder-decoder network predicts facial expressions. FaceFormer uses an autoregressive transformer~\citep{vaswani2017attention} for facial expressions, and AMUSE employs a conditional latent diffusion model~\citep{Rombach_2022_CVPR}. In summary, while EMAGE and TalkSHOW both use VQ-VAE, EMAGE leverages dual training paths (masked gesture recognition and audio-conditioned gesture generation with a switchable cross-attention layer) to effectively merge body hints and audio features and disentangle gesture decoding. In contrast, TalkSHOW trains face and body components separately, autoregressively predicting body and hand motion while incorporating facial expressions from the face decoder. Meanwhile, AMUSE is specially trained for emotional motion generation; since it focuses solely on emotional gesticulation without facial animation, we complement it with FaceFormer for full-body animation sequences.

For dialogue, we generate template responses to scenario-based questions. The text is then fed into a TTS model, which generates speech with appropriate intonation. These intonations drive the emotional arousal-related gestures produced by all models, ensuring alignment between speech and gestures. We use PlayHT TTS~\citep{playht-no-date} to generate emotional speech given text inputs. For a given script, speech is generated with a storytelling narrative style for an adult male, featuring neutral tempo and loudness. Once the models have produced their outputs, GPU acceleration is used to render the meshes in Blender. We incorporate the body shape \(\shape\) parameter and import the \textit{.npz} data into Blender through the \smplx addon~\citep{SMPL-X:2019}, which applies a sample albedo texture upon import, as shown in~\cref{fig:objanim}-top.

\paragraph{Real human animation reconstruction.} We also employ a video-based regression model to reconstruct animations from real actor gestures and expressions, allowing us to compare the performance of synthetic animation against real human motion capture. The model processes a driving video of a real actor and outputs per-frame mesh objects. Specifically, we use PIXIE~\citep{PIXIE:2021} to estimate $\pose$, $\expression$, and gender-specific shape $\shape$ and $\albedo$, while DECA~\citep{Feng:SIGGRAPH:2021} extracts high-fidelity 3D facial displacements. For the reconstruction-based animation, we record an actor responding to scenario-based questions while another individual poses the questions. Video frames are extracted and processed by PIXIE and DECA to obtain geometry, $\albedo$, and lighting information. The audio from the original video is used to synchronize lip movements with the spoken words. Detailed shaded textures, including 3D displacements, are applied by mapping UV textures onto the 3D body mesh on a per-frame basis. Each frame is then exported as a Wavefront OBJ file with shaded textures via PyTorch3D~\citep{ravi2020accelerating}. Finally, using Blender's Geometry Nodes editor, we generate instances of objects from a collection and place them on points derived from the mesh, animating the mesh sequences with the geometry node modifier, as shown in~\cref{fig:objanim}-bottom. All animations share the same outdoor environment background. For inference, we use the default model hyperparameters provided by the original implementations of all methods: EMAGE, TalkSHOW, AMUSE, FaceFormer, PIXIE, and DECA. All input audio was sampled at 16 kHz. We used Blender 3.4 along with the built-in \textit{VR Scene Inspection} add-on for VR streaming. The SMPL-X Blender add-on (v1.1) was used, along with the SMPL-X mesh, textures, and UV map (v1.1, NPZ+PKL format).

\section{User Study}
\label{eval}

\subsection{Research Questions}
We address the following research questions for animations representing two emotional arousal categories:

\begin{itemize}
    \item \textbf{RQ1 (Perceived Animation Realism):} \textit{“Which generative method demonstrates the highest perceived realism during a social interaction?”}
    
    \item \textbf{RQ2 (Perceived Animation Naturalness):} \textit{“Which generative method demonstrates the highest naturalness in terms of facial expressions and bodily gestures?”}
    
    \item \textbf{RQ3 (Perceived Animation Enjoyment):} \textit{“Do the methods influence the perceived level of enjoyment?”}
    
    \item \textbf{RQ4 (Perceived Interaction Quality):} \textit{“Do the methods show differences in the quality of experienced interaction?”}
    
    \item \textbf{RQ5 (Perceived Animation Diversity):} \textit{“Can participants perceive motion diversity between two virtual character animations of the same speech utterance with neutral emotion, presented side by side?”}

    \item \textbf{RQ6 (Perceived Animation Emotion):} \textit{“Can participants correctly identify the arousal level in the generated animation that the model was given as input?”}
\end{itemize}

\subsection{Participants}

We recruited 48 participants (28 males, 20 females) aged 19–48 (\(\mu\) = 26.71, SD = 5.30) via internal channels at the local University. When asked about their recent experiences with virtual environments, 70.8\% reported playing videogames in the past 12 months, and their previous enjoyment with VR experiences varied as follows: ``below average'' (6.25\%), ``average'' (33.3\%), ``good'' (37.5\%), and ``very good'' (22.9\%). All participants were recruited through an internal email system and received a gift card as compensation. The study conformed to the Declaration of Helsinki and was approved by the local ethical committee.

\subsection{Experiment Conditions}\label{subsec:scenario_design}

We conducted a within-subject experiment with two independent variables: method (EMAGE, TalkSHOW, PIXIE+DECA, and
AMUSE+FaceFormer) and scenario (Happy Emotion Animation (HEA), Neutral Emotion Animation (NEA), and Animation Diversity (DV)). The HEA and NEA scenarios involve interactions with an agent displaying happiness and neutral animations, respectively. The DV scenario employs two different PyTorch noise seeds to generate distinct animations of two agents performing the same speech utterance with neutral emotion. In PyTorch, setting a fixed random seed helps control sources of randomness—allowing repeated executions on the same platform and device to produce identical outputs—, which lets us opt into deterministic implementations for certain operations. In the HEA and NEA scenarios, participants engage in one short conversation, whereas in the DV scenario they participate in two conversations. To systematically test method effects, we combined the four animation sources—three generative models and one based on a real human performance—with the three scenarios, yielding twelve experimental conditions. This design enables us to compare the effectiveness of the generative models in expressing emotionally expressive animation both between themselves and against the baseline (PIXIE+DECA) by measuring user perceptions during interaction with the virtual character. The ordering of conditions per participant was counterbalanced using a Latin Square design. The scenario design follows principles from~\citep{ExpressivenessFrontier}. Specific frames from all scenarios are shown in~\cref{fig:objanim}.

\paragraph{Happy Emotion Animation (HEA):} Participants engage in a short conversation where the agent expresses happiness. The prompt is ``\textit{Past accomplishment}'', and the agent responds with ``\textit{I engineered an AI-driven healthcare diagnostic tool, enhancing medical professionals' capabilities for rapid and accurate disease identification and treatment}'', accompanied by consistent gestures and facial expressions generated by each method. This pre-generated response and motion are produced by the system described in~\cref{subsec:genmodels} and configured to convey high arousal (happiness).

    \paragraph{Neutral Emotion Animation (NEA):} This scenario mirrors the HEA condition, but conveying mid arousal instead (neutral emotion). The prompt is ``\textit{Way to relax}'', and the response is ``\textit{I escape to a secluded garden, where the rustle of leaves and blooming flowers ease my mind}''.

    \paragraph{Animation Diversity (DV):} In this scenario, participants encounter two agents under the prompt ``\textit{Christmas plans}''. Both agents respond with ``\textit{This Christmas, I'm eager to create handmade decorations and share the festive spirit with those around me}'', each displaying motion-diverse body gestures and facial expressions generated from neutral emotion input, and presented side by side.

\begin{table*}[t]
\centering
\caption{\normalsize \textbf{Questions used in the perceptual study across VR conditions.}
The table lists each condition (Cond.), the corresponding survey questions, the primary source of the item, and the construct it measures.
Abbreviations for the measures: APR – Animation-Perceived Realism; ANF – Animation-Naturalness of Facial Expressions; ANB – Animation-Naturalness of Body Gestures; AE – Animation Enjoyment; IQ – Interaction Quality; EAR – Animation Emotional Arousal Recognition; AD – Perceived Animation Diversity; SP – Social Presence; OF – Open-ended Feedback; XR - Extended Reality (VR/AR/MR). Cross media represents any media platform—video, XR, or other formats.}

\label{tab:quest}
\small                          %
\setlength{\tabcolsep}{3pt}     %
\begin{tabular}{%
  p{0.05\linewidth}  %
  p{0.40\linewidth}  %
  p{0.19\linewidth}  %
  p{0.15\linewidth}  %
  p{0.14\linewidth}  %
}
\hline
\textbf{Cond.} & \textbf{Questions} & \textbf{Primary \newline Reference} & \textbf{Measures Assessed} & \textbf{Intended Applicability} \\
\hline
Pre &
Did you play any video game in the past 12 months? \newline
How was your experience with virtual environments? \newline
Write positive and negative experiences for these interactions. &
\citep{ExpressivenessFrontier} \newline  \newline \citep{ExpressivenessFrontier} \newline  \newline \citep{ExpressivenessFrontier}  & \vspace{18pt} Exposure & \vspace{18pt} Cross media \\
\hline
HEA &
I felt like I was interacting with a real person. \newline
Their facial expressions looked natural. \newline
Their body movements looked natural. \newline
Rate your enjoyment during the interaction. \newline
Was the interaction warm and comfortable? \newline
What was your perceived emotion of the avatar’s animation? &
\citep{10.1162/105474603322761270} \newline \citep{ExpressivenessFrontier} \newline
\citep{ExpressivenessFrontier} \newline
\citep{Rogers2021RealisticMA}  \newline
\citep{Rogers2021RealisticMA}  \newline
\citep{10.1162/105474603322761270}
& APR/ SP \newline 
ANF \newline 
ANB \newline 
AE \newline 
IQ \newline 
EAR

& XR \newline
Cross media \newline
Cross media \newline
Cross media \newline
Cross media \newline
Emotional interactions \\
\hline
NEA &
I felt like I was interacting with a real person. \newline
Their facial expressions looked natural. \newline
Their body movements looked natural. \newline
Rate your enjoyment during the interaction. \newline
Was the interaction warm and comfortable? \newline
What was your perceived emotion of the avatar’s animation? &
\citep{10.1162/105474603322761270} \newline \citep{ExpressivenessFrontier} \newline
\citep{ExpressivenessFrontier} \newline
\citep{Rogers2021RealisticMA}  \newline
\citep{Rogers2021RealisticMA}  \newline
\citep{10.1162/105474603322761270} & APR/ SP\newline 
ANF \newline 
ANB \newline 
AE \newline 
IQ \newline 
EAR  & XR \newline
Cross media \newline
Cross media \newline
Cross media \newline
Cross media \newline
Emotional interactions  \\
\hline
DV &
Do you perceive diversity in the two animations presented? &
\citep{CONLEY20181059,gestture} & AD & Cross media \\
\hline
Post &
Did you feel the strongest sense of closeness to your conversational partner? \newline
Did the interaction with the avatar feel mostly like a real person? \newline
The avatar’s facial expressions looked the most natural/realistic. \newline
The avatar’s body movements looked the most natural/realistic. \newline
Provide qualitative feedback about your overall VR interaction experience. &
\citep{10.1162/105474603322761270} \newline \newline 
\citep{10.1162/105474603322761270} \newline \newline
\citep{ExpressivenessFrontier} \newline \newline
\citep{ExpressivenessFrontier} \newline \newline
\citep{ExpressivenessFrontier} 
& SP \newline \newline
SP \newline \newline
ANF \newline \newline
ANB \newline \newline
Feedback

& XR \newline \newline
 XR \newline \newline
 Cross media \newline \newline
  Cross media \newline \newline
 Cross media \\
\hline
\end{tabular}
\end{table*}

\subsection{Survey}  

We used a 21-item questionnaire to gauge how each experimental condition influenced perception, social presence, and interaction quality. Three items collected demographics and prior VR exposure, while twelve items—split evenly between Happy and Neutral arousal blocks—assessed perceived realism (from the Networked Minds Social Presence Inventory~\citep{10.1162/105474603322761270}), facial- and body-naturalness (adapted from~\cite{ExpressivenessFrontier}), interaction quality~\citep{Rogers2021RealisticMA}, emotional arousal level~\citep{10.1162/105474603322761270}, and animation diversity~\citep{CONLEY20181059,gestture}. Six additional post-study items, also adapted from~\cite{ExpressivenessFrontier}, captured overall realism, interaction quality, face- and body-naturalness, diversity, and open-ended feedback. All conditions used five-point Likert items, except for perceived emotional arousal, which had three levels (high, medium, low), and the diversity item, which was a binary choice. Some prompts were slightly reworded to match the scope of our study. In~\cref{tab:quest} we provide a complete list of all questions, their primary sources, the subjective metrics they assess, and their intended applicability.

\subsection{Apparatus}\label{apparatus}  
We used an HTC VIVE Pro 2 Head-Mounted Display (90 FPS, 120° FOV, 2448×2448 resolution per eye) with integrated headphones. Two SteamVR 2.0 base stations tracked participants’ positions. The virtual environment was created in Blender 3.4 with OpenXR-based SteamVR integration. The 30 FPS animation was played at 90 Hz in the VR headset using frame duplication, running on a desktop computer with an Intel i9-13900K CPU, 64GB RAM, and an NVIDIA RTX A6000 GPU. To ensure synchronized facial expressions and gestures despite method latency (see~\cref{subsec:inference}), speech and animations were pre-generated before the experiment and then streamed and rendered in real-time during user interaction.

\subsection{Procedure}\label{subsec:experimental_design}  
Participants received an introduction to the study and provided written consent. Once sat down and wearing the headset, they were greeted by a virtual character positioned $1.5 m$ away, allowing them to position themselves comfortably for eye contact. We kept the interpersonal distance and outdoor scene constant, in an effort to eliminate confounding effects of proxemics and place illusion. They then removed the headset to complete a pre-experiment survey. Next, participants experienced the twelve conditions (four methods $\times$ three scenarios) in a counterbalanced order, one trial at a time. Before each trial, they were shown a paper with the conversation prompt and then wore the headset to interact with the virtual character. After each trial, they removed the headset to complete a condition-specific survey, before moving on to the next trial. Upon finishing all scenarios, participants completed a post-experiment survey.

\subsection{Data Analysis}  
Because the collected data did not satisfy the assumption of normality, we employed the aligned rank transform (ART), a non-parametric method suitable for factorial analyses~\citep{10.1145/1978942.1978963}. Specifically, we used an ART ANOVA for all statistical tests and applied Bonferroni corrections for pairwise comparisons.

\section{Results}
\label{res}

\paragraph{\textbf{Perceived animation realism.}}  
We found that the methods did not significantly influence realism: EMAGE (\(Md = 2, IQR = 2\)), TalkSHOW (\(Md = 3, IQR = 2\)), PIXIE+DECA (\(Md = 3, IQR = 2\)), and AMUSE+FaceFormer (\(Md = 3, IQR = 2\)). This result was confirmed by a non-significant main effect for methods (\(F(3, 141) = 1.5, p = 0.2, \eta^2 = 0.03\)). However, we discovered that the happy emotion condition (\(Md = 3, IQR = 2\)) yielded higher realism ratings than the neutral emotion condition (\(Md = 2.5, IQR = 2\)), as supported by a statistically significant main effect for emotion (\(F(1, 47) = 11.5, p < 0.001, \eta^2 = 0.2\)). Finally, no statistically significant interaction effect was observed for methods \(\times\) emotion (\(F(3, 141) = 1.6, p = 0.17, \eta^2 = 0.03\)).

\paragraph{\textbf{Perceived animation naturalness of facial expressions.}}  
PIXIE+DECA (\(Md = 3, IQR = 2\)) resulted in higher ratings for the naturalness of facial expressions compared to EMAGE (\(Md = 2, IQR = 1\)), TalkSHOW (\(Md = 3, IQR = 2\)), and FaceFormer (\(Md = 2, IQR = 1\)). This was confirmed by a statistically significant main effect for methods (\(F(3, 141) = 3.3, p = 0.02, \eta^2 = 0.07\)). Pairwise comparisons showed significant differences between EMAGE and PIXIE+DECA (\(p = 0.01\)), but not among the other pairs (\(p > 0.05\)). No significant differences were found between the happy emotion (\(Md = 3, IQR = 1\)) and neutral emotion (\(Md = 2, IQR = 1\)) conditions (\(F(1, 47) = 1.49, p = 0.22, \eta^2 = 0.03\)). However, a statistically significant interaction effect for methods \(\times\) emotion was observed (\(F(3, 141) = 4.1, p = 0.007, \eta^2 = 0.08\)). Pairwise comparisons revealed significant differences between the neutral emotion condition in EMAGE and PIXIE+DECA (\(p = 0.01\)), and between TalkSHOW happy emotion and EMAGE neutral emotion (\(p = 0.0238\)); the remaining comparisons were not significant (\(p > 0.05\)).

\paragraph{\textbf{Perceived animation naturalness of body gestures.}}  
We found that methods did not significantly affect the naturalness of bodily movements: EMAGE (\(Md = 3, IQR = 2\)), TalkSHOW (\(Md = 3, IQR = 2\)), PIXIE (\(Md = 3, IQR = 2\)), and AMUSE (\(Md = 3, IQR = 2\)). This was confirmed by a non-significant main effect for methods (\(F(3, 141) = 1.3, p = 0.26, \eta^2 = 0.03\)). However, the happy emotion condition (\(Md = 3, IQR = 2\)) resulted in higher naturalness ratings than the neutral emotion condition (\(Md = 3, IQR = 2\)), a difference supported by a statistically significant main effect for emotion (\(F(1, 47) = 6.4, p = 0.01, \eta^2 = 0.12\)). No significant interaction effect was observed for methods \(\times\) emotion (\(F(1, 141) = 1.57, p = 0.19, \eta^2 = 0.03\)).

\paragraph{\textbf{Perceived animation enjoyment.}}  
We found that the methods did not significantly influence enjoyment levels: EMAGE (\(Md = 3, IQR = 2\)), TalkSHOW (\(Md = 3, IQR = 2\)), PIXIE+DECA (\(Md = 3, IQR = 1.25\)), and AMUSE+FaceFormer (\(Md = 3, IQR = 2\)). Similarly, there was no significant difference between the happy emotion (\(Md = 3, IQR = 2\)) and neutral emotion (\(Md = 3, IQR = 2\)) conditions. These findings were supported by non-significant main effects for methods (\(F(3, 141) = 2.4, p = 0.06, \eta^2 = 0.05\)) and emotion (\(F(1, 47) = 2.6, p = 0.11, \eta^2 = 0.05\)). Additionally, no statistically significant interaction effect was found for methods \(\times\) emotion (\(F(3, 141) = 1.05, p = 0.36, \eta^2 = 0.022\)).

\begin{figure*}[t]
    \centering
    \includegraphics[width=1.04\linewidth]{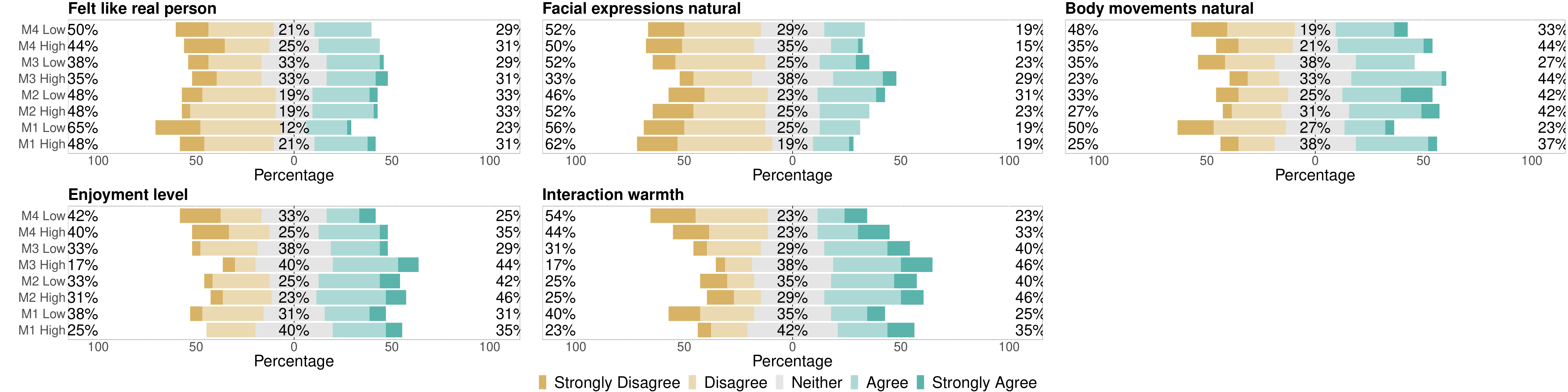}
    \caption{\textbf{Summary of likert scale results.}
    Summary of Likert scale ratings for Animation Realism (avatar felt like a real person), Animation Naturalness (facial expressions; body movements), Animation Enjoyment, and Interaction Quality (interaction warmth). For brevity, we denote EMAGE, TalkSHOW, PIXIE+DECA, and AMUSE+FaceFormer as M1, M2, M3, and M4, respectively, and use ``High" and ``Low" to represent happy and neutral emotions.}
    \label{likert}
\end{figure*}

\paragraph{\textbf{Perceived interaction quality.}}  
TalkSHOW (\(Md = 3, IQR = 1.25\)) resulted in higher ratings for interaction quality compared to EMAGE (\(Md = 2, IQR = 1\)), PIXIE+DECA (\(Md = 3, IQR = 1\)), and AMUSE+FaceFormer (\(Md = 3, IQR = 2\)). This difference was supported by a statistically significant main effect for methods (\(F(3, 141) = 4.2, p < 0.01, \eta^2 = 0.08\)). Pairwise comparisons indicated significant differences between TalkSHOW and AMUSE+FaceFormer (\(p = 0.027\)), while the other comparisons were not significant (\(p > 0.05\)). No significant differences were observed between the happy (\(Md = 3, IQR = 2\)) and neutral (\(Md = 3, IQR = 2\)) emotion conditions (\(F(1, 47) = 4, p = 0.051, \eta^2 = 0.07\)). Furthermore, no significant interaction effect was found for methods \(\times\) emotion (\(F(3, 141) = 1.57, p = 0.2, \eta^2 = 0.03\)). A summary of the Likert scale results for realism, facial expressions, bodily movements, enjoyment, and interaction quality is shown in~\cref{likert}.

\begin{table}[t]
  \centering
  \caption{\textbf{Arousal recognition rates by method and sequence.}
  For each animation method and sequence (H = high-arousal for happiness, N = mid-arousal for neutral), participants’ percent correct identification in the three response categories.}
  \label{tab:arousal_rates}
  \footnotesize
  \setlength{\tabcolsep}{3.5pt}
  \renewcommand{\arraystretch}{1.1}
  \begin{tabularx}{\columnwidth}{@{}c c *{3}{>{\centering\arraybackslash}X}@{}}
    \toprule
    \textbf{Method} & \textbf{Seq.} & \textbf{High} & \textbf{Mid} & \textbf{Low} \\
    \midrule
    \multirow{2}{*}{\textbf{EMAGE}} &
      H & \cellcolor{SoftBlue}55.5  & 35.2 & \cellcolor{SoftPurple}9.3 \\
      & N & 10.8 & \cellcolor{SoftBlue}72.2 & \cellcolor{SoftPurple}17.0 \\
    \midrule
    \multirow{2}{*}{\textbf{TalkSHOW}} &
      H & \cellcolor{SoftBlue}56.0 & 34.8 & \cellcolor{SoftPurple}9.2 \\
      & N & 5.8 & \cellcolor{SoftBlue}78.4 & \cellcolor{SoftPurple}15.8 \\
    \midrule
    \multirow{2}{*}{\textbf{PIXIE+DECA}} &
      H & \cellcolor{SoftBlue}61.5 & 30.4 & \cellcolor{SoftPurple}8.1 \\
      & N & 2.2 & \cellcolor{SoftBlue}89.6 & \cellcolor{SoftPurple}8.2 \\
    \midrule
    \multirow{2}{*}{\textbf{AMUSE+FaceFormer}} &
      H & \cellcolor{SoftBlue}70.8 & 23.1 & \cellcolor{SoftPurple}6.1 \\
      & N & 6.8 & \cellcolor{SoftBlue}74.4 & \cellcolor{SoftPurple}18.8 \\
    \bottomrule
  \end{tabularx}
\end{table}

\begin{table}[t]
\centering
\caption{\textbf{Emotion recognition accuracy for happy and neutral animations.}
Each row shows the softmax probability distribution (\%) over the eight Ekman emotions for a given method and sequence.
N, H, A, S, C, Su, F, and D represent Neutral, Happiness, Anger, Sadness, Contempt, Surprise, Fear, and Disgust, respectively.
See~\cref{sec:Arousaleffects} for discussion.}
\label{tab:emotion_recognition}
\scriptsize
\setlength{\tabcolsep}{2.6pt}
\renewcommand{\arraystretch}{1.15}

\begin{tabularx}{\columnwidth}{@{}c c *{8}{>{\centering\arraybackslash}X}@{}}
\toprule
\textbf{Method} & \textbf{Seq.} & \textbf{N} & \textbf{H} & \textbf{A} & \textbf{S} & \textbf{C} & \textbf{Su} & \textbf{F} & \textbf{D} \\
\midrule
\multirow{2}{*}{\textbf{EMAGE}}
 & H & 8.5  & \cellcolor{SoftGold}19.5 & 9.2  & \cellcolor{SoftPurple}30.2 & 7.8 & 8.1 & 8.0 & 8.7 \\
 & N & \cellcolor{SoftGold}20.0 & 8.2  & 8.5  & \cellcolor{SoftPurple}31.0 & 9.0 & 8.0 & 8.0 & 7.3 \\
\midrule
\multirow{2}{*}{\textbf{TalkSHOW}}
 & H & \cellcolor{SoftPurple}17.5 & \cellcolor{SoftGold}28.0 & 10.8 & 8.3 & 9.6 & 8.2 & 10.5 & 7.1 \\
 & N & \cellcolor{SoftGold}30.0 & 8.5  & 8.2  & 8.8 & 9.1 & 9.0 & \cellcolor{SoftPurple}18.0 & 8.4 \\
\midrule
\multirow{2}{*}{\textbf{PIXIE+DECA}}
 & H & 6.8 & \cellcolor{SoftBlue}44.0 & \cellcolor{SoftPurple}18.7 & 6.4 & 6.6 & 6.5 & 5.2 & 5.8 \\
 & N & \cellcolor{SoftBlue}44.2 & 6.9 & 5.4 & 8.0 & 5.3 & 6.5 & \cellcolor{SoftPurple}16.8 & 6.9 \\
\midrule
\multirow{2}{*}{\textbf{AMUSE+FaceFormer}}
 & H & 3.2 & \cellcolor{SoftBlue}56.0 & 3.6 & 3.5 & 3.7 & \cellcolor{SoftPurple}18.9 & 5.5 & 5.6 \\
 & N & \cellcolor{SoftBlue}54.3 & 5.0 & 3.5 & \cellcolor{SoftPurple}19.0 & 4.8 & 2.4 & 5.7 & 5.3 \\
\bottomrule
\end{tabularx}
\end{table}

\paragraph{\textbf{Animation emotional arousal recognition.}}  
As the last question in the six-item survey, participants rated the animation’s arousal for both HEA and NEA conditions. After being told to judge the perceived emotional arousal of each clip, they chose one of three options: high, medium, or low arousal. Overall, participants correctly identified high‐arousal clips 60.94\% of the time and mid‐arousal clips 78.65\% of the time.  By method, EMAGE had a recognition percentage of 55.5\% on high and 72.2\% on mid, TalkSHOW 56.0\% and 78.4\%, PIXIE + DECA 61.5\% and 89.58\%, and AMUSE + FaceFormer 70.83\% and 74.4\%, respectively.  Thus, AMUSE + FaceFormer led in high‐arousal recognition, while PIXIE + DECA excelled at mid‐arousal detection. The detailed confusion matrix for two stimulus levels (high and mid arousal) across three response options (high, mid, low) is shown in~\cref{tab:arousal_rates}. Correct identifications are highlighted in \colorbox{SoftBlue}{blue}, while any confusions in which a high- or mid-arousal stimulus was classified as low arousal are shaded in \colorbox{SoftPurple}{violet}.

To further analyze arousal recognition, we used a deep learning–based motion extractor~\citep{petrovich21actor,chhatre2023emotional} trained on motion capture data to predict one of eight emotion classes (an extended Ekman-style eight-class taxonomy: neutral, happy, angry, sad, contempt, surprise, fear, disgust). We present the predicted emotion recognition probabilities in~\cref{tab:emotion_recognition}, where the best-performing methods’ happy and neutral sequences are highlighted in \colorbox{SoftBlue}{blue} and second best in \colorbox{SoftGold}{yellow}, while the emotions with which the method is confused are highlighted in \colorbox{SoftPurple}{violet}. 

\paragraph{\textbf{Animation diversity.}}  
Participants were asked to judge whether two side-by-side virtual-character animations—generated from distinct initial conditions as described in~\cref{subsec:scenario_design}—appeared diverse. Because this diversity item was a binary choice, we did not subject it to statistical analysis. AMUSE+FaceFormer was rated most effective, with 95.8\% of participants perceiving diversity. In contrast, EMAGE received the lowest ratings, with 70.8\% reporting perceived diversity and 18.8\% indicating no diversity. Both TalkSHOW and PIXIE+DECA received 79.2\% of participants reporting perceived diversity. To complement our statistical analysis, we computed the Euclidean distance (2-norm) between joints on the SMPL-X axis angles, yielding diversity scores of 2.5336 for EMAGE, 2.0777 for TalkSHOW, and 2.9360 for AMUSE+FaceFormer; PIXIE+DECA shows no diversity due to its deterministic reconstruction approach.

\paragraph{\textbf{Post-experiment feedback: Perceived closeness and realism.}}  
Participants evaluated their experiences using 5-point Likert scale responses regarding closeness, perceived realism, and the naturalness of facial expressions and bodily movements. The post‐study items were collected only once per participant—after all methods had been experienced—we treat these four measures as overall user impressions rather than method‐specific comparisons. Accordingly, we report only descriptive statistics and do not perform factorial tests. Post-study ratings yielded a median sense of closeness ($Md = 2, IQR = 1$), agent realism ($Md = 3, IQR = 2$), facial-expression naturalness ($Md = 3, IQR = 1$), and body-gesture naturalness ($Md = 3, IQR = 2$), indicating an overall mildly positive perception of the virtual character’s social presence and animation quality. EMAGE and TalkSHOW received the lowest ratings, with 28 and 30 participants, respectively, rating closeness as ``A little.'' PIXIE+DECA performed best, with 24 participants reporting ``Quite a bit'' of closeness and 27 finding the agent realistic. PIXIE+DECA also scored highest for natural facial expressions, with 23 participants rating them as ``Quite a bit.'' AMUSE+FaceFormer received more balanced feedback, with 22 participants finding the agent realistic and 23 rating the bodily movements as natural.

\section{Discussion}
\label{dis}

\subsection{Scenario Design}

Our scenarios focus on everyday conversations, each associated with an internal emotional state. For example, a relaxation topic corresponds to a neutral emotional stance, while a past achievements topic evokes a happy emotional state. This setup explores how varying emotional cues affect behavior and perception. In a ``passion" scenario, audio and gestures convey energetic or happy expressions, whereas in a ``relaxation" scenario, they are more subdued. We then evaluate the extent to which the methods can generate distinguishable emotional levels. Regarding animation diversity, we measure how much gesture variation is acceptable before the virtual character’s identity appears inconsistent for the same speech utterance, as if a participant was interacting with a different entity within the same scenario.

\subsection{Emotional 3D Animation}\label{sec:Arousaleffects}

Our findings show that the emotion category significantly affects animation realism. Happy emotion animations with energetic gestures were perceived as more realistic than neutral emotion animations, indicating that high-arousal, happy expressions have a stronger social presence during an interaction across all methods (RQ1). In terms of facial expression naturalness, PIXIE+DECA outperformed the other methods—especially in neutral emotion scenarios—demonstrating a superior ability to capture subtle facial cues. Additionally, emotional interaction revealed that PIXIE+DECA consistently performed better (particularly compared to EMAGE in neutral conditions), primarily due to DECA's robust capture of 3D facial displacements, which enhances the base expressions predicted by PIXIE; no speech-driven face animation method is comparable to the reconstruction-based real human facial expressions compatible with SMPL-X meshes (RQ2-face). For body movement naturalness, emotion again played a key role: happy emotion movements were rated more natural (RQ2-body), mirroring the results for animation realism (RQ1). While enjoyment levels were similar across all methods (RQ3), TalkSHOW outperformed the others in interaction quality—especially when compared to EMAGE and AMUSE+FaceFormer—suggesting that TalkSHOW’s output may support a stronger interactive connection with users (RQ4).

Survey data shows that 60.94\% of participants correctly identified the happy emotion condition, while 78.65\% correctly recognized the neutral emotion condition, indicating that mid-arousal gestures were easier to identify (RQ6). PIXIE+DECA achieved the highest accuracy (89.58\%) for neutral emotion, whereas AMUSE+FaceFormer performed best for happy emotion (70.83\%), demonstrating that AMUSE+FaceFormer animations are easier to recognize for happy emotion compared to other methods. EMAGE exhibited balanced accuracy for both conditions, while TalkSHOW and PIXIE+DECA showed a trend toward more accurate mid-arousal identification.

The effectiveness of each model in generating distinguishable emotional levels depends on its architecture and processing approach (see \cref{tab:comparison_methods}). All methods are generative and probabilistic, but differ in their preprocessing approaches; EMAGE and AMUSE include unique processing steps, whereas TalkSHOW uses standard inputs without specialized preprocessing.

\paragraph{EMAGE and AMUSE}. Extract disentangled latent representations for speech content, emotion, style, and rhythm. These robust representations allow for better alignment with arousal cues rather than merely producing varied animations.

\paragraph{PIXIE}. Operates purely on video input, reconstructing realistic animation directly from a human actor’s performance. Although this can yield high-quality results, it relies on the actor’s expressiveness and does not create new gestures.

Using this statistical deep learning emotion recognition metric for motion sequences, we observe that AMUSE+FaceFormer and PIXIE+DECA demonstrate the highest emotion recognition scores, with AMUSE+FaceFormer achieving 56\% accuracy for happy emotion and 54.3\% for neutral emotion. Specifically, AMUSE+FaceFormer predictions confused Happy with Surprise and Neutral with Sad; PIXIE+DECA confused Happy with Angry and Neutral with Fear. In contrast, TalkSHOW and EMAGE demonstrate lower emotion recognition accuracy, with TalkSHOW confusing Happy with Neutral and Neutral with Fear, and EMAGE showing high confusion with Sad in both sequences. These quantitative findings align with our user study data, indicating that PIXIE+DECA excels at capturing high-quality animations, although depending on the actor’s performance, whereas audio-based methods can independently generate synthetic animations with disentangled emotion and content, producing more clearly distinguishable gesture arousal—with the speech-driven method AMUSE+FaceFormer showing the highest accuracy.

\subsection{Animation Diversity}

The perceived animation diversity varied significantly across models, with AMUSE+FaceFormer standing out—95.8\% of participants noticed diverse animations. In contrast, EMAGE scored lowest at 70.8\%, while 79.2\% of participants observed diversity with the other models. Animation diversity is essential for crowd animations and extended interactions, where varied gestures, movements, and contexts create engaging, lifelike experiences; it also applies to both speech-driven and idle animations, which are key to maintaining natural behavior in virtual characters. Additionally, quantitative measurements of animation diversity—computed as the Euclidean (2-norm) distance between SMPL-X joint axis angles—show that AMUSE+FaceFormer has the highest 2-norm, followed by EMAGE and TalkSHOW. These findings reinforce our statistical analysis, confirming that greater animation diversity enhances perceived interaction quality and supports RQ5.

\subsection{Inference Times}
\label{subsec:inference}
The inference times for producing 10-second animation sequences were as follows: EMAGE required $0.827 s$; TalkSHOW, $20.29 s$; PIXIE+DECA, $412.63 s$; and AMUSE+FaceFormer totaled $8.561 s$ ($2.557 s$ for body animation, plus $5.337 s$ for face animation). EMAGE is the fastest, making it particularly efficient for real-time or near real-time applications. AMUSE+FaceFormer strikes a balance between speed and complexity, being faster than TalkSHOW but slightly slower than EMAGE, while PIXIE+DECA is by far the slowest due to the complexity of video-based animation reconstruction.

\subsection{Design Recommendations}

In our evaluation, we compared state-of-the-art speech-driven 3D emotional animation generation methods to examine their strengths and weaknesses, as well as how they shape user perception in VR. By comparing these generative approaches with the reconstruction of a real actor, we also investigated how closely current methods can replicate real human body and facial expressions. We note that marker-based motion capture yields higher-quality real-actor motion than our reconstructed animations. Based on our user study results, we note the following design recommendations:

\paragraph{Emotional modeling.}  
    While speech-to-animation methods often focus on lip-sync and body gestures, explicit emotion modeling is frequently overlooked. As shown in~\cref{tab:emotion_recognition}, all animation methods (EMAGE, TalkSHOW, AMUSE, FaceFormer) have considerable scope for improvement in emotion recognition (RQ6). Although AMUSE, which is trained on explicit audio emotion and person identity modeling for gesture generation, shows the best relative accuracy (56.0\% for happiness and 54.3\% for neutral), there remains a gap. Emotions are often confused with other emotional animation sequences across models. AMUSE achieves emotion modeling by disentangling driving speech into content, emotion, and style; however, its approach is limited to a single categorical emotion per sequence. Exploring multiple concurrent emotions in an animation sequence is a promising direction for future research. Finally, scenario context may affect emotion perception: participants inferred emotions not only from gestures but also from the spoken content or from how convincingly the actor’s performance was rendered (for the video reconstruction method).

    \paragraph{Animation generation for emotional states with lower arousal.}  
    In terms of animation realism (RQ1) and naturalness (RQ2 for both body and face), we observed that animations representing high-arousal, happy emotions consistently received higher ratings than those for neutral, low-arousal states. The generative models generally perform better when generating pronounced expressions compared to subtle, idle movements. This is likely because the motion datasets used for training typically include expressive sequences rather than calm, idle ones. Incorporating mocap datasets focused on calm or idle motions, such as breathing-based movement, could help models generalize to less exaggerated animations.

    \paragraph{Joint modeling of facial expressions and body gestures.}  
    Among the evaluated methods, EMAGE jointly trains face and body, whereas TalkSHOW trains them separately within the same framework, and AMUSE does not address facial expressions. Even with joint training in EMAGE, no method achieved high ratings for facial expression naturalness (RQ2-face). This highlights the challenge of simultaneously learning both expression parameters and body gestures, largely due to the differences in data representations (face data uses 100 SMPL-X expression parameters, while body data is based on joint rotations in the world coordinate system). More robust data preprocessing and unified parameterization are needed to effectively train a single model for both full-body and facial animations.

    \paragraph{Dyadic interaction feedback.}  
    All generative methods exhibited similarly low performance in animation enjoyment (RQ3) and interaction quality (RQ4), although TalkSHOW showed relatively higher interaction quality. This suggests that current evaluations, which often rely solely on statistical metrics, do not fully capture the user-centric experience in immersive interaction settings. Incorporating user-centered evaluation into the feedback mechanism is crucial, as it ensures that the generated animations effectively convey the intended emotion and enhance user enjoyment and interaction.

\section{Limitations and Future Work}
\label{lim}

Our evaluation represents a useful first step but has several limitations that suggest valid directions for future research—namely, addressing latency issues, exploring the full spectrum of emotion categories, varying VR hardware setups, incorporating additional behavioral measures, and benchmarking against video-based reconstruction—all of which are detailed below:

\paragraph{Latency and turn-taking.}  
    Our evaluation employs a modular approach using several large deep generative models. For instance, AMUSE—a latent diffusion model with 440 million parameters—generates temporal SMPL-X motion parameters but suffers from slow inference due to extensive denoising steps and high GPU memory requirements (e.g., RTX A6000 with 48 GB). Similarly, the face generation models in TalkSHOW and FaceFormer, which are based on Transformer autoregressive architectures, experience longer inference times due to their sequential design. As these latencies limit real-time applications, we pre-generate speech and animations and stream them in real time for single-turn conversations, as noted in~\cref{apparatus}. Supporting multi-turn conversations, however, would require a fully real-time setup with no pre-generation, where the speech responses and corresponding animations must be generated and streamed simultaneously in VR. As noted in~\cref{subsec:inference}, EMAGE is currently the most suitable method for real-time interaction, achieving the lowest latency at 0.827s. To account for this, we introduce a fixed 5-second idle movement period between turns, where the agent adopts a neutral, forward-facing stance. In a fully real-time multi-turn conversation, idle motion and wait times would need to dynamically adapt to the length of the participant's response. To the best of our knowledge, no currently available system can generate idle body motion and trigger speech and animation generation in real time upon detecting the end of a user’s reply. Achieving this would require integrating speech-to-speech models with real-time animation generation, followed by synchronized playback of speech and gestures—an avenue we identify as future work. Such a system would also require careful consideration of hardware, as computing demands are expected to remain high.

    \paragraph{Emotion categories.}  
  Emotions can be described as discrete classes or as points in a continuous affective space (see~\cref{subsubsec:emo_in_socint}).  
  To keep the study tractable, we restricted our experiment to two conditions—Happy (high arousal) and Neutral (mid arousal)—which provided clear initial insights while avoiding an exponential growth in the number of possible condition combinations. Extending the protocol to cover additional emotions across the full arousal–valence spectrum of the circumplex model is an important goal for future work.

    \paragraph{VR apparatus.}  
    VR streaming and rendering are compute-intensive and heavily hardware-dependent. As described in~\ref{apparatus}, we mitigated these challenges by splitting the interaction into two phases: real-time streaming for interaction and precomputed full-body animation rendering, with other components generated in advance. Future improvements in VR hardware for rendering and streaming will further alleviate these limitations.
    
    \paragraph{Video-based reconstruction.}  
    Although our video-based reconstruction method shows promising per-frame quality, its temporal coherence is lower. Frame-by-frame pose estimation, when played back at 30 FPS, leads to jittery animations (see supplementary video). Despite our initial expectation that video-based reconstruction of the real animation would yield the best performance across enjoyment (RQ3) and interaction quality (RQ4), our user study revealed that reconstruction methods did not excel in these areas, even though facial animation (RQ2-face) was enhanced by DECA-based face displacements. Future studies should explore reconstruction methods that improve temporal coherence and pose estimation for smoother animations at the desired frame rate.

    \paragraph{Additional behavioral measures.} Research on human-like behaviour in virtual agents remains in its early stages. Our study is a useful first step, but a richer evaluation is needed—particularly on metrics such as eye-gaze patterns and task-completion time. Concepts central to believability and presence—including co-presence, plausibility, place illusion, the uncanny valley for interactive agents, and both subjective and inter-subjective symmetry—also require analysis. In addition, more complex social dynamics (e.g., group interaction and contact behaviour such as self-contact, interpersonal contact, and ground contact) should be examined. Progress will depend on developing stronger generative models and testing them in more sophisticated realistic environments.

\section{Conclusion}
\label{conc}

We present an evaluation of generative models for emotional 3D animation within an immersive VR environment, focusing on user-centric metrics—emotional arousal realism, naturalness, enjoyment, diversity, and interaction quality—in a real-time human–virtual character interaction scenario through a user study ($N=48$). In this study, we systematically examined perceived emotional quality across three state-of-the-art speech-driven 3D animation methods and compared them to a real human reconstruction-based animation under two emotional conditions: happiness (high arousal) and neutral (mid arousal). Participants recognized emotions more accurately for generative methods that explicitly modeled animation emotions. User study data showed that generative models performed well for high-arousal emotion but struggled with subtle arousal emotion. Although reconstruction-based animations received higher ratings for facial expression quality, all generative methods exhibited lower ratings for animation enjoyment and interaction quality, highlighting the importance of incorporating user-centric evaluations into generative animation model development. All methods demonstrated acceptable animation diversity; however, differing inference times among generative methods, along with VR rendering latency, posed limitations. Lastly, while the video-based reconstruction method (compatible with SMPL-X meshes) produced high-quality frame-level animations from driving videos, it lacked temporal coherence, leading to suboptimal performance in user ratings of animation enjoyment and interaction quality. Overall, these findings highlight the importance of integrating user-centric evaluations into the development of generative models to produce virtual animated agents that outperform rule-based and teleoperated techniques. Hence, we believe that evaluating models solely on technical metrics during development is insufficient to ensure that the animations convey the perceptual details we want end users to experience in conversational scenarios.

\section{Acknowledgements}
\label{Acknowledgements}

We thank Peiyang Zheng and Julian Magnus Ley for support with the user-study technical setup, and Tairan Yin for proofreading and feedback. This project was funded by the EU Horizon 2020 program under the Marie Skłodowska-Curie grant agreement No. 860768 (CLIPE) and by the Swedish Research Council through grant 2020-05187.

{
    \small
    \bibliographystyle{ieeenat_fullname}
    \bibliography{main}
}
\bigskip
{\noindent \large \bf {APPENDIX}}\\

\renewcommand{\thefigure}{A.\arabic{figure}} 
\setcounter{figure}{0} 
\renewcommand{\thetable}{A.\arabic{table}}
\setcounter{table}{0} 

\appendix

The supplementary material provides an overview of the supplementary video content in~\cref{sec:supmatvid}. It also provides survey details in~\cref{sec:QA} and further setup details in~\cref{sec:expset}.

\section{Supplementary Video}
\label{sec:supmatvid}

The supplementary video\footnote{\href{https://play.kth.se/media/Evaluation+of+Generative+Models+for+Emotional+3D+Animation+Generation+in+VR/0_q1e7393r}{Supplementary video (KTH Play)}} provides an overview of the generated gestures and interactions, including the following:

\begin{enumerate}
    \item An explanation of the state of the art in VR-based interaction and gesture generation using generative models.
    \item An example interaction displaying the four methods, with one example condition for each: HA, LA, CG, and DV.
    \item A comparison of all four methods in terms of gesture quality, focusing solely on the HA condition.
    \item A side-by-side comparison of an exemplar method in both HA and LA conditions to illustrate the differences between the two arousal levels.
    \item A summary of the results addressing the three research questions, followed by an overall conclusion.
\end{enumerate}

\section{Survey}
\label{sec:QA}
\subsection{Survey Questionairae}

The full study for each participant took approximately one hour, though this varied depending on individual interaction times. At the beginning of the session, we conducted an exemplar interaction to familiarize participants with the environment.

\subsection{Participant Qualitative Feedback} 
\label{sec:qualfeed}
Now, we present the participants' selective responses to the qualitative feedback questions below.

\subsubsection{Arousal}
\begin{itemize}
    \item \textbf{P6 (M1)}: ``High sound volume. Much better experience when you can feel the intensity of interaction.''
    \item \textbf{P19 (M4)}: ``The movement made the interaction feel more alive, with gestures being more prominent.''
    \item \textbf{P31 (M1)}: ``Way too much energy, looked a bit like they were over-exaggerating.''
\end{itemize}

\subsubsection{Congruence}
\begin{itemize}
    \item \textbf{P2 (M3)}: ``Both the facial and body expression is more natural this time, and more aligned with the sound.''
    \item \textbf{P17 (M4)}: ``The movement of the avatar seemed to match the conversation very well.''
    \item \textbf{P12 (M2)}: ``Movements were sluggish and unsync with voice cues, which felt odd.''
\end{itemize}

\subsubsection{Diversity}
\begin{itemize}
    \item \textbf{P48 (M3)}: ``The last series of videos had more distinguishable gestures, which kept it engaging.''
\end{itemize}

\subsubsection{Detailed Feedback on Facial and Body Movements}
\begin{itemize}
    \item \textbf{P9 (M1)}: ``Movements were sluggish and unsync with voice cues, which felt odd.''
    \item \textbf{P10 (M2)}: ``I could only perceive the movements of the mouth clearly; facial gestures felt static.''
    \item \textbf{P16 (M1)}: ``Eyes were very static. Lips didn’t seem to move naturally.''
    \item \textbf{P30 (M3)}: ``Movement with mouth was realistic, but the expressions still felt stiff.''
    \item \textbf{P42 (M1)}: ``Strange posture and eye movement, but nice hand gestures.''
\end{itemize}

\paragraph{}Participants provided mixed feedback on arousal, with some appreciating the heightened intensity of gestures (P6, M1; P19, M4), while others found the energy excessive (P31, M1). For congruence, improvements in synchronization were noted (P2, M3; P17, M4), though issues with sluggish or unsynced movements were highlighted (P12, M2). Diversity feedback was positive, with P48 (M3) noting that gesture variety kept the interaction engaging. However, concerns were raised regarding facial movements, with participants mentioning static or unsynced facial gestures (P9, M1; P10, M2) and stiffness in expressions (P16, M1; P30, M3), though hand gestures were generally well received (P42, M1). Overall, while body gestures were generally more natural, facial animations require further refinement.

\begin{figure}[t]
    \centering
    \includegraphics[width=\columnwidth]{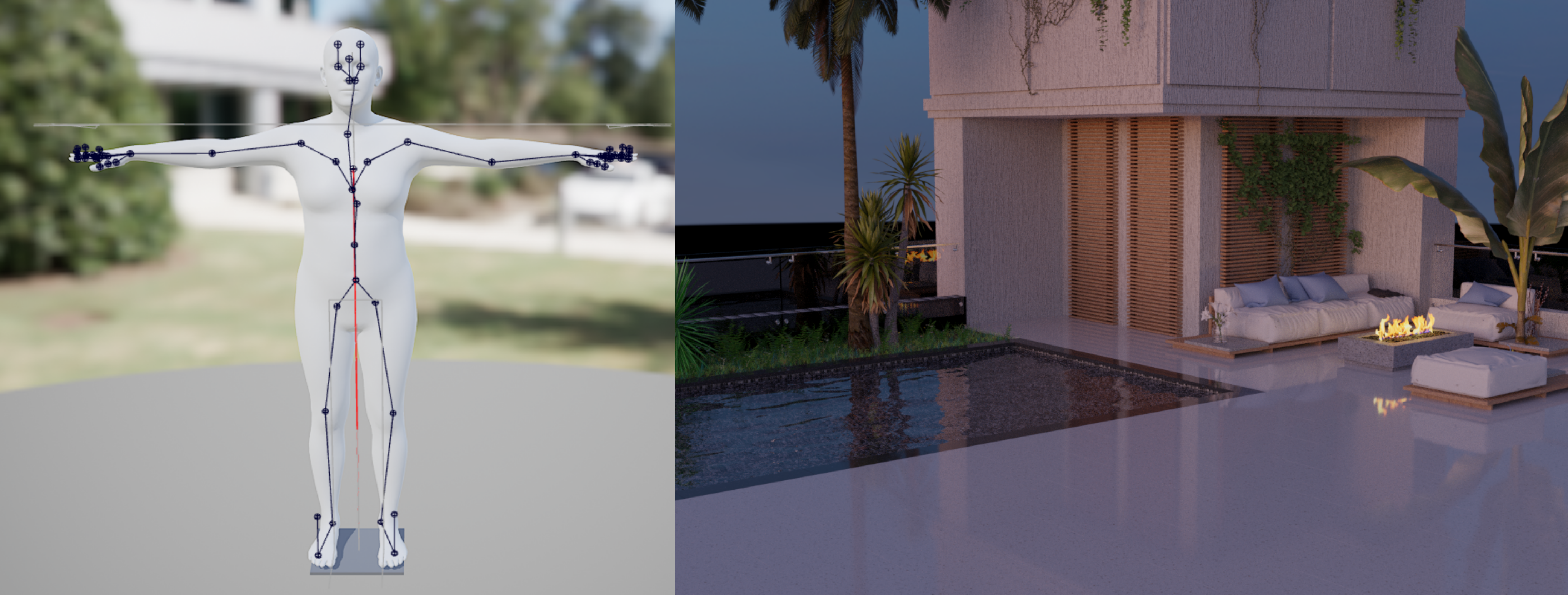}
    \caption{Left: \textbf{SMPL-X joints.} Right: \textbf{Blender render of an outdoor scene.}}
    \label{blenren}
\end{figure}

\begin{table}[t]
\centering
\caption{\textbf{SMPL-X joint names with indices.}}
\label{tab:smplx_joint_names}
\scriptsize
\setlength{\tabcolsep}{3pt}
\renewcommand{\arraystretch}{1.05}

\begin{minipage}{0.49\columnwidth}
\centering
\begin{tabular}{@{}c l@{}}
\toprule
\textbf{Idx} & \textbf{Joint} \\
\midrule
0  & Pelvis \\
1  & Left Hip \\
2  & Right Hip \\
3  & Spine1 \\
4  & Left Knee \\
5  & Right Knee \\
6  & Spine2 \\
7  & Left Ankle \\
8  & Right Ankle \\
9  & Spine3 \\
10 & Left Foot \\
11 & Right Foot \\
12 & Neck \\
13 & Left Collar \\
14 & Right Collar \\
15 & Head \\
16 & Left Shoulder \\
17 & Right Shoulder \\
\bottomrule
\end{tabular}
\end{minipage}\hfill
\begin{minipage}{0.49\columnwidth}
\centering
\begin{tabular}{@{}c l@{}}
\toprule
\textbf{Idx} & \textbf{Joint} \\
\midrule
18 & Left Elbow \\
19 & Right Elbow \\
20 & Left Wrist \\
21 & Right Wrist \\
22 & Jaw \\
23 & Left Eye \\
24 & Right Eye \\
25--27 & Left Index1--3 \\
28--30 & Left Middle1--3 \\
31--33 & Left Pinky1--3 \\
34--36 & Left Ring1--3 \\
37--39 & Left Thumb1--3 \\
40--42 & Right Index1--3 \\
43--45 & Right Middle1--3 \\
46--48 & Right Pinky1--3 \\
49--51 & Right Ring1--3 \\
52--54 & Right Thumb1--3 \\
\bottomrule
\end{tabular}
\end{minipage}
\end{table}

\section{Experimental Setup Details} 
\label{sec:expset}

\subsection{SMPL-X Model}

\Cref{tab:smplx_joint_names} and~\cref{blenren} (Left) shows the SMPL-X jointa and lists their corresponding indices. The mesh typically represents a person with height 1.7 m, weight 60 kg, and 55 number of joints (\textbf{J}).

\subsection{Lighting and Setup in Blender}

This outdoor scene~\cref{blenren} (Right) shows a patio with seating, surrounded by greenery and a pool. The lighting creates a evening ambiance with a fire source near the sofa. The Blender lighting setup uses High Dynamic Range Image (HDRI) for ambient light, while point lights illuminate the fire. The distance between the camera and the chair was consistently maintained at 1.5 meters, ensuring a uniform setup. The only variation was due to participants' different torso heights and preferred viewing angles.

\end{document}